\documentclass[useAMS,usenatbib]{mnras}
\usepackage{graphicx,amssymb,amsmath}
\usepackage{txfonts, stfloats}
\usepackage{xcolor}
\usepackage{hyperref}
\usepackage{xspace}

\usepackage{lipsum}

\usepackage{siunitx}
\usepackage{comment}
\usepackage{soul}
\usepackage{subcaption}
\usepackage{url}

\newcommand{\omrel}{\Omega_{\rm rel}}
\newcommand{\ommat}{\Omega_{\rm m}}
\newcommand{\omrad}{\Omega_\gamma}
\newcommand{\omnu}{\Omega_\nu}
\newcommand{\omlambda}{\Omega_\Lambda}
\newcommand{\omgw}{\Omega_{\rm GW}}
\newcommand{\cosmotherm}{{\tt CosmoTherm}\xspace}

\newcommand{\neff}{N_{\rm eff}}
\newcommand{\Tgw}{\mathcal{T}_{\rm GW}}
\newcommand{\lcdm}{$\Lambda$CDM}

\newcommand{\pot}[2]{#1 \times 10^{#2}}

\newcommand{\Mpc}{{{\rm Mpc}}}

\graphicspath{{./Immagini_high_dpi/}}

\definecolor{light-gray}{gray}{0.70}

\title[Clarifying transfer function approximations]{Clarifying transfer function approximations for the large-scale gravitational wave background in $\mathbf{\Lambda}$CDM}

\begin{document}

\author[Kite et al.]{
Thomas Kite$^{1}$\thanks{E-mail: \href{mailto:thomas.kite@manchester.ac.uk}{thomas.kite@manchester.ac.uk}},
Jens Chluba$^{1}$\thanks{E-mail: \href{mailto:jens.chluba@manchester.ac.uk}{jens.chluba@manchester.ac.uk}}
Andrea Ravenni$^{1}$\thanks{E-mail: \href{mailto:andrea.ravenni@manchester.ac.uk}{andrea.ravenni@manchester.ac.uk}},
and Subodh P. Patil$^{2}$\thanks{E-mail: \href{mailto:patil@lorentz.leidenuniv.nl}{patil@lorentz.leidenuniv.nl}}
\\
$^1$Jodrell Bank Centre for Astrophysics, School of Physics and Astronomy,
The University of Manchester, Manchester, M13 9PL, U.K.
\\
$^2$Instituut-Lorentz for Theoretical Physics, Leiden University, 2333 CA Leiden, The Netherlands
}

\date{\vspace{-0mm}{Accepted 2020 --. Received 2020 --}}

\maketitle

\begin{abstract}
The primordial gravitational wave background (GWB) offers an exciting future avenue of discovery for new physics. Its information content encodes multiple eras in the early Universe's history, corresponding to many orders of magnitude in frequency and physical scale to be measured today.
By numerically solving for the GW transfer functions we provide simple yet accurate formulas describing the average power of the large-scale energy spectrum of the GWB for arbitrary primordial tensor power spectra.
In doing so we can pedagogically explain and clarify previous GWB literature, highlight the important cosmological parameters of various GWB features, and reveal multiple ways in which cancelling conceptual errors can give deceptively accurate results.
The scales considered here are particularly important for CMB probes of the GWB, via $B$-modes and spectral distortions.
In particular, we carefully study the effects of both neutrino damping, and the precise nature of the transition between the radiation-dominated (RD) and matter-dominated (MD) eras.
A byproduct of numerically solving the problem is the ability to study the robustness of common approximations in the literature. Specifically, we show that a numerical treatment is especially important around the RD--MD transition, and for a brief moment of history where neutrino damping occurs during MD.
In passing we also discuss the effects of late acceleration caused by dark energy -- showing that this can be neglected in most practical GWB applications -- and the effects of changing relativistic degrees of freedom on the GWB at very small-scales.
\end{abstract}


\section{Introduction}
\label{sec:introduction}
The detection of the first gravitational wave (GW) \citep{LIGOScientific:firstGW} opened a door to a  novel way of studying the Universe. Decades of studying the light arriving from the cosmos has provided us with modern precision cosmology as we know it, and with some poetic license, we are now able to \textit{hear} the Universe as well as \textit{see} it.

The excitement of this prospect has led to a suite of new upcoming probes (either proposed or under construction) which will listen for GWs in different frequency bands \citep[see][for review]{Campeti:2020xwn}. From lowest to highest frequency GWs we have CMB $B$-mode measurements \citep{Ade:2018gkx, Aghanim:2018eyx}, spectral distortion measurements \citep{Kite:2021GW2SD}, pulsar timing array measurements \citep{Perera:2019sca, Alam2020NG}, and finally direct detection using interferometry \citep{Abbott2020ApJ, Abbott2020II}. Through a combination of all these probes we can construct a comprehensive picture of the symphony of GWs in the Universe, and refine our understanding of fundamental physics in the process.

In this work, we focus on primordial origins of GWs rather than astrophysical sources. Our study therefore relates to searches for a stochastic gravitational wave background (GWB) rather than single isolated events. The exact physics that will be revealed through studying this background is broad and diverse \citep[see][for review]{Caprini:2018mtu}.

The goal of this paper is then twofold: firstly to pedagogically introduce the physics of the GWB to clarify other literature, revealing potential pitfalls in the analytic modelling, and secondly to provide a simple yet accurate analytic description for the mapping between the present-day large-scale GWB energy spectrum and the corresponding primordial tensor power. The latter allows our results to be applied to general inflationary models, making this work particularly relevant to the interpretation of $B$-mode and spectral distortion searches for new physics.

The mapping from underlying physical model to present-day observations requires a detailed understanding of the GW transfer function, for which various solutions have been considered \citep[e.g.,][]{Watanabe2006, Dicus:2004:nuDamp, Caprini:2018mtu}. We expand upon this literature with a numerical treatment of the GWB which accounts for the nuanced cosmological expansion through radiation-dominated (RD) and matter-dominated (MD) eras, the late time accelerated expansion from dark energy (DE) and the non-negligible damping from free-streaming neutrinos. This allows us to give the promised simple fits for the average large-scale GWB energy spectrum in a number of fiducial cosmological scenarios.

Accurately accounting for the transition between RD and MD eras is especially important in calculations of the transfer function for non-standard thermal histories, such as those with epochs of early matter domination frequently encountered in a variety of phenomenological extensions of the standard cosmology \citep{Acharya:2008bk, Acharya:2019pas}, or for scenarios where the primordial GW spectrum is significantly enhanced or modulated, relevant, for instance, in scenarios of primordial black hole formation \citep{Ballesteros:2019hus, Bhattacharya:2021wnk,Green:2020jor, Arbey:2020yzj}. We will discuss how the results of this paper can also be straightforwardly extended to such applications.

This paper is organised as follows: 
in Sect.~\ref{sec:GWB_physics} we qualitatively review the broad range of fundamental physics imprinted on the GWB. 
This will aid the reader in understanding the more quantitative approach in Sect.~\ref{sec:analytic_solutions}, where we analytically solve the equation governing the evolution of GWs in limiting cases. These solutions, although previously considered, will serve to clarify some confusion in the literature about their application.
The numerical method is explained and results shown in Sect.~\ref{sec:numerical_solutions}, focusing on the reliability of the analytic results previously found. One region of parameter space not captured well by existing approximations is the MD--RD transition, which is important for CMB scale probes. Hence in this section we provide simple fits for the large-scale GWB, providing an alternative to the usual analytic approximations.
More general features of the GWB are discussed in Sect.~\ref{sec:GWB_features}, where we demonstrate the principal cosmological dependence of neutrino damping and the main effects of late time acceleration on the GWB. For completeness we include some discussion of changes in the relativistic degrees of freedom and their relevance to SD constraints on GW backgrounds. We point out in this section how a combination of the simple fits and pretabulated data on relativistic degrees of freedom can accurately model the spectrum to within $\sim 5\%$ on all scales.
Finally we summarise and conclude in Sect.~\ref{sec:conclusion}.

\section{Physics content of the GWB}
\label{sec:GWB_physics}
The study of cosmological perturbation theory explains the evolution of perturbations on the otherwise smooth expanding FLRW background, and is the foundation for much of modern cosmology \citep{Ma1995}. For detailed explanation and derivations with details about GWs see \cite{WeinbergBook}, but we summarise the essential steps here. Perturbatively small terms are added to both the metric $g_{\mu\nu}$ and the stress energy tensor $T_{\mu\nu}$, which can then be equated through Einstein's field equations. Three fundamental types of perturbations emerge from this calculation: scalars, transverse vectors and spatial transverse traceless tensors. The latter are what we also understand as GWs. These waves couple to the corresponding spatial transverse traceless tensor component within $T_{\mu\nu}$, the anisotropic stress of the medium, $\Pi$, which provides a source term that can damp the GWB.

This last point is quite important, as typically speaking the particle species in the primordial plasma do not carry considerable anisotropic stresses: tightly-coupled fluids rapidly isotropize and are dominated by their densities and velocities, after which comes a period of free streaming dominated solely by velocity\footnote{For a more general analysis that interpolates between the kinetic and hydrodynamic regimes, incorporating ambient matter interactions, see for instance \citep{Baym:2017xvh, Flauger:2017ged, Miron-Granese:2020hyq, Zarei:2021dpb}.}. Only a brief intermediate phase therefore leads to a non-negligible anisotropic stress that can interact with and damp the GWB. The dominant damping effects therefore arise from the GWs themselves sourcing the anisotropic stress in the medium, which will lead to an integro-differential equation that we solve numerically.

A subdominant contribution to the damping is added by the cosmic photon field. At early times the photon fluid inherits enough energy from the GWB to produce a noteworthy distortion to the blackbody spectrum \citep{Chluba2015}, but with no discernible effect on the GWB. The GWB scales most affected by photons are $k\simeq 10^{-2}\,\Mpc^{-1}$, amounting to a $14\%$ reduction in the amplitude squared according to the work of \cite{saikawa:2018}. However, we note that at these scales it is both possible and necessary to model the photon decoupling with the full Boltzmann equation, rather than using a modified version of the damping term [e.g., Eq.~\eqref{eq:damping_integrand} below], which contains several simplifying assumptions. The damping effect of photons will not significantly change the results of this paper, and a full detailed treatment is left to future work.

The neutrino, on the other hand, has a considerable damping effect over a large set of scales. Previous studies show that the neutrino field will damp the GWB amplitude squared by $\simeq 35.6\%$ \citep{Weinberg:2004:nuDamp, Dicus:2004:nuDamp} at scales $k\gtrsim 1/500\,\Mpc^{-1}$. The damping effect arising from neutrinos will be investigated below, verifying and generalising on these previous studies. We note that it is conceivable to treat the neutrino field with the same level of sophistication as the photon field: understanding how inherited energy from the GWB will distort the otherwise thermal distribution of neutrino momenta, and modelling a gradual decoupling of the particles through full Boltzmann hierarchies. However, also this program is beyond the scope of this paper.

The bottom line then is that within the standard thermal history of the Universe, the GWB is mostly free from the surrounding plasma, only receiving small predictable damping effects from free streaming neutrinos. The rest of the information encoded in the GWB therefore comes from the state of the Universe at the time of horizon crossing for each frequency, after which simple propagation occurs. This is, in fact, the double-edged sword of GW cosmology: a feeble interaction that simultaneously makes a clean and powerful probe of almost the entirety of cosmological history, but which also makes for an incredibly difficult detection at present time. A detection is a sufficiently monumental task that glimpsing the GWB has become the aspiration of many scientific teams, with a diverse set of probes.

One important state of the Universe's history cleanly imprinted as a GWB feature is the precise moment that relativistic particle species no longer dominate the universal expansion, giving way to a matter dominated era. Since GWs have a different evolution in each of the eras, there is a predictable change in shape of the energy spectrum (see Sect.~\ref{subsec:simple_fits}). One goal of this paper is to elucidate this transition in order to facilitate comparison between early and late Universe probes of the GWB.

To model the moment of this transition it is important to cleanly separate the cosmic inventory into relativistic and non-relativistic particles. This usually equates to distinguishing massive and massless species, but some subtleties arise when considering neutrinos. We now know from data on neutrino oscillations \citep{neutrino:oscillations:1998, neutrino:oscillations:2001, neutrino:oscillations:2002} to expect massive neutrinos, albeit with masses limited to sub-eV scales \citep{Planck2018params,Aker:2021gma}. The concordance model in Cosmology therefore still treats these as massless entities in most applications. This is often sufficient since the sum of neutrino masses is predicted to be sufficiently small that the early-universe dynamics will resemble that of massless particles, even if at least two of the neutrino species must be non-relativistic today \citep{Lesgourgues:2006nd}.

In this paper, we therefore carefully distinguish the photon energy density, $\omrad=\pot{5.42}{-5}\,[T_0/2.7255\,{\rm K}]^4\,[h/0.675]^{-2}$, from the total relativistic energy density
\begin{equation}
    \omrel = \omrad + \omnu = \left(1+\neff \left[\frac{7}{8}\right]\left[\frac{4}{11}\right]^{4/3}
    \right)\omrad,
\end{equation}
which includes the neutrino energy density $\omnu$. The number of relativistic degrees of freedom, $\neff$, parameterizes the extra massless degrees of freedom relative to the photons. The factor of $7/8$ arises due to the differences in particle statistics (Bose-Einstein or Fermi-Dirac), while the factor $(4/11)^{4/3}$ relates to the energy release during electron-positron annihilation. In this paper we assume the standard model expectation value of $\neff=3.046$ \citep{Mangano2005, deSalas:neutrino:2016}, which in turn gives $\omrel=\pot{9.18}{-5}$ today. This distinction between the photon field and the full relativistic cosmic inventory has been ambiguous or neglected in some literature, leading to additional confusion around the exact moment of RD--MD transition \citep[e.g. see discussion in sect.~5.2 in][]{Caprini:2018mtu}. As previously mentioned, resolving this disparity is important for accurate comparison between the largest scale CMB $B$-Modes and spectral distortion measurements, and constitute one driving motivation for this work.

One more energy component needs to be included to complete the cosmic inventory: the cosmological constant or dark energy\footnote{For the purposes of this paper, $\omlambda$ will be referred to as dark energy and cosmological constant interchangeably -- only dark energy with $w=-1$ is considered.} $\omlambda$. Despite being the dominant form of energy today, it makes up a tiny fraction of the Universe's content at primordial times. The expected effect of this component is only small changes on the largest physical scales, which can be verified numerically (Sect.~\ref{subsec:late_acceleration}). A more notable difference from the late-time acceleration is the change in the age of the Universe, which complicates the application of analytic solutions, as we clarify here.

The physics discussed thus far is all needed to accurately model the GWB down to scales of $k\simeq 10^3 \,\Mpc^{-1}$. Beyond these scales the spectral features arise from changes in the number of relativistic degrees of freedom, $g_*$, as originally discussed in \cite{Watanabe2006}, generalised by \cite{Boyle2008a}, and recently solved to high precision by \cite{saikawa:2018}. These changes in the energy budget, arising from the cooling effect of the universal expansion, cause small temporary changes in the expansion rate, which is imprinted on the GWB from the moment of horizon crossing. We will briefly discuss the importance of these effect on spectral distortion constraints, leaving the details of the physics to the aforementioned papers.

\section{Analytic GW solutions}
\label{sec:analytic_solutions}
The equation of motion governing the evolution of a GW, derived from cosmological perturbation theory, is given by \citep{Weinberg:2004:nuDamp,Watanabe2006,Boyle2008a}
\begin{equation}
    \partial^2_\eta h_k^\lambdaup+2\frac{a'}{a}\partial_\eta h_k^\lambdaup+k^2 h_k^\lambdaup = 16\pi Ga^2\Pi^\lambdaup,
    \label{eq:master_gw_eq}
\end{equation}
where $h_k^\lambdaup(\eta)$ is the amplitude of the gravitational wave at wavenumber $k$ for each polarization $\lambdaup = +,\times$, and $\Pi^\lambdaup(k,\eta)$ is the anisotropic stress of the surrounding primordial plasma, both as a function of wavenumber $k$ and conformal time $\eta$. Primes denote derivatives respect to conformal time, but we keep some explicit derivatives for clarity later where we will change coordinates.
The amplitude of a physical GW can be written as the product of a transfer function with some initial amplitude $h_k^\lambdaup(\eta) = h_k^{\lambdaup,\text{prim}}\Tgw(k, \eta)$, and as such we have $\Tgw(k,0)=1$. This decomposition of transfer function and initial condition helpfully separates the statistical from the deterministic, as well as distinguishing the inflationary from the post-reheating dynamics.

A primary goal of this paper is to give simple yet precise estimates for the energy density of the GWB, which measured relative to the critical density is given by
\begin{equation}
    \omgw(k) = \frac{\rho_{\rm GW}}{\rho_{\rm c}}(k) = \frac{\mathcal{P}_T(k)}{12a^2H^2}[\Tgw'(k)]^2.
    \label{eq:energy_density}
\end{equation}
Here, the primordial tensor power spectrum
\begin{equation}
    \mathcal{P}_T(k)
    =
    \frac{2 k^3}{2 \pi^2}
    \sum_\lambdaup
    \langle |h_k^{\lambdaup,\text{prim}}|^2 \rangle
\end{equation}
encodes the statistical properties of the initial conditions via an ensemble average\footnote{We have followed the convention of \cite{Watanabe2006} and \cite{saikawa:2018}, which can be expressed in terms of other conventions by noting the normalisation of polarisation tensors in the latter reference, between Eq.~(2.4) and Eq.~(2.5).}. For many applications the energy density is the essential quantity one needs to know, since any experiment measuring the GWB is sensitive to its energy density at a given time and scale/frequency. It is clear from Eq.~\eqref{eq:energy_density} that fundamental link between the primordial $\mathcal{P}_T$ and $\omgw$ at any other time is the transfer function $\Tgw$, which we study in detail next.

\subsection{Transfer function}
\label{subsec:transfer_functions}
As previously discussed, a key feature in the GWB is a distinctive \textit{bend} on physical scales corresponding to the transition between the radiation-dominated\footnote{We remind the reader that despite the misnomer we include relativistic neutrinos here.} (RD) and matter-dominated (MD) eras of the Universe's history. To understand this effect it is instructive to first ignore both the contribution of DE and the effects of damping - the former being negligible and the latter being an unnecessary complication to describe the physics of the transition. Solving the Friedman Equations in this limit we have
\begin{subequations}
\label{eq:MD-RD-expansion}
\begin{gather}
\eta=2\frac{\sqrt{a\ommat + \omrel}-\sqrt{\omrel}}{H_0\ommat},
\label{eq:eta_RDMD}\\
a=\frac{1}{4}\eta^2H_0^2\ommat + \eta H_0 \sqrt{\omrel},
\label{eq:scale_factor_RDMD}\\
\frac{a'}{a} = aH = aH_0\sqrt{\Omega_{\rm m} a^{-3} + \Omega_{\rm rel} a^{-4}}.
\label{eq:hubble_RDMD}
\end{gather}
\end{subequations}
Using these expressions one can find
\begin{subequations}
\label{eq:simple_aprime_a}
\begin{gather}
    \frac{a'}{a} = \frac{1}{\eta} + \frac{1}{\eta + \eta_*} = \frac{1}{\eta_*}\left(\frac{1}{\xi} + \frac{1}{1+\xi}\right),\\
    \eta_* = 1/k_* = 4\sqrt{\omrel}/H_0\ommat.
\end{gather}
\end{subequations}
The characteristic time-scale defined here is $\eta_*=540.44\,\Mpc$ for up-to-date cosmological parameters from \cite{Planck2018over}. With this time-scale, the dimensionless quantities $\xi=\eta/\eta_*$ and $\kappa=k/k_*=k\eta_*$ naturally emerge. Using these variables is advantageous for various reasons, but most notably it adds a degree of invariance in considering different cosmologies. Note the commonly appearing term $\kappa\xi=k\eta$, which helps in matching to common approximations in the literature. Another common time-scale for RD--MD equality is $a_{\rm eq}=\omrel/\ommat$, defined simply as the time in which energy densities of the respective components matched\footnote{It is often unclear which time-scale an author uses, and as such we will keep a strict convention here. The $\Tgw$ approximations by \cite{Watanabe2006}, which we discuss shortly, give the correct limiting cases using $\eta_*$ as defined in this work.}.

By incorporating this change of variables to the differential equation we find an elegant form
\begin{equation}
    \partial^2_\xi\Tgw+2\left(\frac{1}{\xi}+\frac{1}{1+\xi}\right)\partial_\xi\Tgw+\kappa^2\Tgw\approx0.
    \label{eq:GW_eom_dimless}
\end{equation}
The characteristic time-scale used here can be further motivated by noticing that it is the time that balances the two contributions to Eq.~\eqref{eq:scale_factor_RDMD}, showing it is closely related to the balance of matter and radiation.

Using Eq.~\eqref{eq:GW_eom_dimless} it is possible to study the evolution of GWs far into both RD ($\xi\ll 1$) and MD ($\xi\gg 1$). In each limiting case we obtain
\begin{equation}
    2\left(\frac{1}{\xi}+\frac{1}{1+\xi}\right) \longrightarrow
    \begin{cases}
    2/\xi \,\,\,\,\,\text{for RD}\\
    4/\xi \,\,\,\,\,\text{for MD}
    \end{cases}
    ,
    \label{eq:xi_limits}
\end{equation}
which offer simple solutions to Eq.~\eqref{eq:GW_eom_dimless} in terms of spherical Bessel functions which we summarise here:
\begin{subequations}
\label{eq:free_transfer}
\begin{gather}
    \Tgw^{\rm RD} = A j_0(\kappa\xi) - B y_0(\kappa\xi),\\
    \Tgw^{\rm MD} = \frac{3}{\kappa\xi}\left[C j_1(\kappa\xi) - D y_1(\kappa\xi)\right],
\end{gather}
\end{subequations}
with derivatives
\begin{subequations}
\label{eq:free_transfer_prime}
\begin{gather}
    \Tgw^{\prime\,\,{\rm RD}} = -k \left[A j_1(\kappa\xi) - B y_1(\kappa\xi)\right],\\
    \Tgw^{\prime\,\,{\rm MD}} = -\frac{3k}{\kappa\xi}\left[C j_2(\kappa\xi) - D y_2(\kappa\xi)\right].
\end{gather}
\end{subequations}
Here $A$, $B$, $C$ and $D$ are constants determined from initial conditions and matching conditions which we discuss below. Note that the derivatives here are still with respect to conformal time, which yields factors of $k$. The terms involving spherical Bessel functions of the first kind, $j_n$, are constant at early times, and have been scaled here such that $A=C=1$ gives an early time normalisation to unity. Spherical Bessel functions of the second kind, $y_n$, are the decaying modes.

The solutions given above are each valid deep into each regime, but we have yet to discuss the transition between them. Note first of all that MD scales ($\kappa\ll 1$) simply stay constant in the RD era, since for those modes we have $\kappa\xi\ll 1$. On the contrary we must be careful with the RD scales ($\kappa\gg 1$) during the MD era, since these modes have already had time to evolve and decay by that time. An approximation for this matching process is performed by \cite{Watanabe2006} (henceforth WK06), where by assuming an instantaneous transition one can solve
\begin{subequations}
\begin{gather}
    \Tgw^{\rm RD}\bigg\rvert_{\xi=1} = \,\, \Tgw^{\rm MD}\bigg\rvert_{\xi=1},\\
    \Tgw^{\prime\,\,{\rm RD}}\bigg\rvert_{\xi=1} = \,\, \Tgw^{\prime\,\,{\rm MD}}\bigg\rvert_{\xi=1},
\end{gather}
\end{subequations}
which gives a functional form to the constants previously defined:
\begin{subequations}
\label{eq:normalisation}
\begin{gather}
A = 1,\\
B = 0,\\
C(\kappa) = \frac{1}{2} - \frac{\cos(2\kappa)}{6} + \frac{\sin(2\kappa)}{3\kappa},\\
D(\kappa) = -\frac{1}{3\kappa} + \frac{\kappa}{3} + \frac{\cos(2\kappa)}{3\kappa} + \frac{\sin(2\kappa)}{6}.
\end{gather}
\end{subequations}
This matches the equations given by WK06 once accounting for different variable conventions. A noteworthy difference in convention is the lack a step function that enforces $C\rightarrow1$ and $D\rightarrow0$ for $\kappa\ll 1$, which is already the natural tendency of the functions as they are written here. Notice that the constant mode solution for the RD scales will excite a decaying mode in the MD era, and thus $D$ cannot be ignored even if $B$ has been.

\subsection{Energy spectrum}
\label{subsec:energy_spectrum}
By using the analytic forms derived in Sect.~\ref{subsec:transfer_functions} the expected limits of the energy spectrum Eq.~\eqref{eq:energy_density} can be derived. Recalling that we are only interested in the power spectrum normalised energy density as seen today ($\eta_0\gg\eta_*$), we can take the limit in the MD era:
\begin{equation}
\label{eq:MD_energy_limit}
\begin{aligned}
    \frac{\Omega_{\rm GW}}{\mathcal{P}_T}\bigg\rvert_{\xi_0} &= \frac{1}{12 H_0^2}\left[\Tgw^{\prime\,\,{\rm MD}}\right]^2\\
    &= \frac{1}{12 H_0^2} \frac{9}{\eta_*^2\xi_0^2} \left[C j_2(\kappa\xi_0) - D y_2(\kappa\xi_0)\right]^2,
\end{aligned}
\end{equation}
from which the high and low $\kappa$ limits can be derived. In both limits however we note that any realistic probe of the GWB will have sensitivity on scales much smaller than those crossing horizon in recent times ($k \gg 1/\eta_0$). This statement leads to $\kappa\xi_0 \gg 1$ expansions, for which we have \citep{Watanabe2006}
\begin{subequations}
\label{eq:today_energy_density}
\begin{align}
    j_n(x) \approx \frac{\sin(x-n\pi/2)}{x} &\,\,\,\,\,\,\text{for}\,\,x\gg 1,\\
    \big\langle j_n(x)^2 \big\rangle \approx \left(\frac{1}{2}\right)\frac{1}{x^2} &\,\,\,\,\,\,\text{for}\,\,x\gg 1,
\end{align}
\end{subequations}
where angle brackets indicate averages over an oscillation, leading to an explicit\footnote{For clarity we will keep this convention of explicit $1/2$ throughout the paper.} factor of $1/2$.

For fixed $\xi_0$ and \textit{large} $\kappa$, the dominant term in the expansion of Eq.~\eqref{eq:today_energy_density} will have a linear term $\kappa/3 \subset D(k)$ combined with  $-\cos(\kappa\xi_0)/\kappa\xi_0 \subset y_2(\kappa\xi_0)$. This gives a flat (albeit oscillating) spectrum to high frequencies:
\begin{equation}
\label{eq:energy_high_kappa}
\begin{aligned}
    \bigg\langle\frac{\Omega_{\rm GW}}{\mathcal{P}_T}\bigg\rvert_{\xi_0}\bigg\rangle &\,\stackrel{\stackrel{\kappa\gg 1}{\downarrow}}{\approx}
    \frac{9}{12 H_0^2\eta_*^2\xi_0^2} \bigg\langle\left[ -\frac{\kappa}{3}\frac{\cos(\kappa\xi_0)}{\kappa\xi_0}  \right]^2\bigg\rangle\\
    &\stackrel{\stackrel{\kappa\xi_0\gg 1}{\downarrow}}{\approx}\left(\frac{1}{2}\right)\frac{\eta_*^2}{12 H_0^2\eta_0^4}\\ 
    &\,\,\,=\,\,\, \left(\frac{1}{2}\right)\frac{\omrel}{12}.
\end{aligned}
\end{equation}
Note however in the final equality we have used a value of $\eta_0$ derived from Eq.~\eqref{eq:scale_factor_RDMD}, by setting $a_0=1$. This may appear problematic, since dark energy dominates the expansion from $a\gtrsim 3/4$, and thus changes the age of the Universe. The analytic approximations derived here however were derived explicitly in a Universe without DE, and should not be used in conjunction with DE-modified values of $\eta_0$. This \textit{cancellation of errors} is vindicated by the numerical solutions (see Sect.~\ref{subsec:late_acceleration}).

To investigate the behaviour at \textit{low} $\kappa$, we start with $C(\kappa)\rightarrow1$, $D(\kappa)\rightarrow0$, and again apply the subhorizon condition $\kappa\xi_0 \gg 1$. This suggests the dominant term being $-\sin(\kappa\xi)/\kappa\xi\subset j_2(\kappa\xi)$. A similar calculation to above gives
\begin{equation}
\label{eq:energy_low_kappa}
\begin{aligned}
    \bigg\langle\frac{\Omega_{\rm GW}}{\mathcal{P}_T}\bigg\rvert_{\xi_0}\bigg\rangle &\,\stackrel{\stackrel{\kappa\ll 1}{\downarrow}}{\approx}\frac{1}{12 H_0^2} \frac{9}{\eta_*^2\xi_0^2} \bigg\langle\left[-\frac{\sin(\kappa\xi_0)}{\kappa\xi_0}\right]^2\bigg\rangle\\
    &\stackrel{\stackrel{\kappa\xi_0\gg 1}{\downarrow}}{\approx}\left(\frac{1}{2}\right) \frac{\omrel}{12} \frac{9}{\kappa^2}.\\
\end{aligned}
\end{equation}
These results will be used in Sect.~\ref{subsec:simple_fits} to choose a functional form for an envelope fit to the data, and in turn verify the accuracy of the numerical calculations.
\subsection{Anisotropic stress}
\label{subsec:anisotropic_stress}
\begin{figure}
\centering 
\includegraphics[width=\columnwidth]{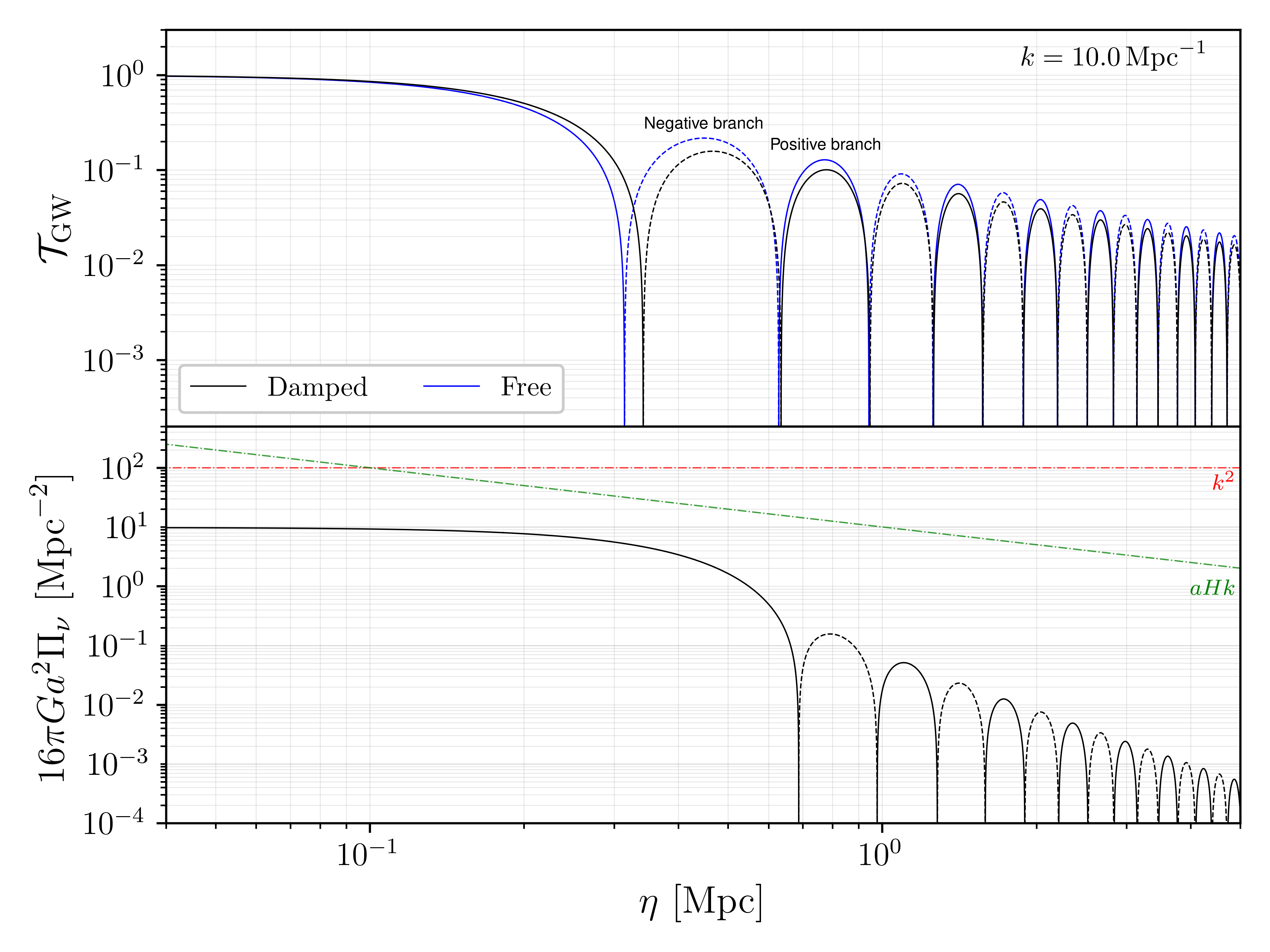}
\caption{A figure showing an example of the neutrino damping term and its effect on the transfer function at $k=10\,\Mpc^{-1}$. The top panel reveals small phase shifts accompany a drop in amplitude of $\Tgw$ around the time of horizon crossing. Dashed lines indicate negative branches of the oscillating function. The lower panel shows the damping term $16\pi Ga^2\Pi_\nu$. Dotted lines show other terms in the differential equation associated with the wavenumber and Hubble expansion.}
\label{fig:damping_example}
\end{figure}
We previously discussed that some particles will contribute to the anisotropic stress of the medium, and constitute damping terms to the GW solution. These stresses are excited by the propagation of the GW itself, and hence makes Eq.~\eqref{eq:master_gw_eq} an integro-differential equation, requiring a more careful treatment than the previous section. We do not derive any analytic solutions here, but instead quote the results of \cite{Dicus:2004:nuDamp} (henceforth DR04). We give the form of the damping integral here both for completeness, and to motivate a numerical approach to solving the problem, as described in Sect.~\ref{sec:numerical_solutions}.

Explicitly evaluating the RHS of Eq.~\eqref{eq:master_gw_eq} for the case of neutrinos gives \citep{Weinberg:2004:nuDamp}
\begin{subequations}
\begin{align}
\label{eq:damping_integrand}
    16\pi Ga^2\Pi_\nu^\lambdaup&=-24 f_\nu\left(\frac{a'}{a}\right)^2\int^\eta_{\eta_{\nu}}
    K(k[\eta-\bar{\eta}])\,
    \Tgw^{\prime}(\bar{\eta})\,
    h_k^{\lambdaup,\text{prim}}\,
    {\rm d}\bar{\eta}, \\
    f_\nu &= \frac{\Omega_\nu}{\Omega_\gamma+\Omega_\nu}\frac{1}{1+a/a_{\rm eq}} = \frac{f_{\nu,0}}{1+a/a_{\rm eq}},\\[1mm]
    K(x) &= \frac{1}{16}\int_{-1}^1 (1-s^2)^2 {\rm e}^{isx}{\rm d}s =
    \frac{j_2(x)}{x^2}
    \nonumber \\[1mm] 
    &=
    \frac{3\sin(x)}{x^5} - \frac{3\cos(x)}{x^4} - \frac{\sin(x)}{x^3},
    \label{eq:damping_terms}
\end{align}
\end{subequations}
where $\eta_\nu$ is the time at which neutrinos decouple, corresponding to a temperature of $\simeq 2$ MeV \citep[e.g.,][]{Jeong2014}.

An example of this damping term is shown in Fig.~\ref{fig:damping_example}, where it is seen that at the time of horizon crossing there is a significant damping of the wave followed by a period of regular propagation, albeit at a lower overall amplitude. The dotted lines show approximate amplitudes for other terms in Eq.~\eqref{eq:master_gw_eq}, revealing that the damping is subdominant, and comparable only at horizon crossing. 

We note in passing that more general particle interactions in the collision time approximation contribute an additional exponential suppression inside the integrand of Eq.~\eqref{eq:damping_integrand} of the form $\exp\,[-\int^\eta_{\bar\eta} \frac{d\eta'}{\tau_c(\eta')}]$,  where $\tau_c$ is the average time between particle collisions \citep{Baym:2017xvh}, making manifest that tightly coupled particles rapidly isotropize and suppress anisotropic stresses, whereas free streaming particles, for which $\tau_c \to \infty$ reduces to Eq.~\eqref{eq:damping_integrand}. We also note that \cite{saikawa:2018} use a modified expression for the neutrino energy fraction $f_{\nu}$ which includes energy inherited from $e^+e^-$ annihilation. This leads to a slightly greater damping effect at scales of $k\sim\pot{3}{4}\,\Mpc^{-1}$, quickly adopting the same asymptotic limit as found in this paper (See Fig.~\ref{fig:GW_gstar}).

Within the RD era, the damped transfer functions are given by DR04 in the form of a series sum of spherical bessel functions:
\begin{equation}
    \Tgw(k\eta) = \sum_{n=0}^{\infty} a_{2n} j_{2n}(k\eta).
\end{equation}
Although in principle this sum has infinitely many terms, in practice only a few are needed. We will take this series with the 7 coefficients provided by DR04 as a benchmark in the RD era, but differences are expected as the universe becomes more matter dominated.

\section{Numerical solutions}
\label{sec:numerical_solutions}
Evaluating the damped solution (see Sect.~\ref{subsec:anisotropic_stress}) has all the usual difficulties of an integro-differential equation: it involves an integral over the history of the GW's own velocity, and it cannot be easily pretabulated since the integrand depends on the upper limit of the integral itself. In this work we use an iterative method to achieve the solution to within some desired accuracy: the first iteration of the method assumes no damping [therefore solving Eq.~\eqref{eq:GW_eom_dimless}] to achieve an initial \textit{guess} $\Tgw^{(0)}$. Each subsequent iteration calculates $\Tgw^{(N)}$ by inserting $\Tgw^{(N-1)}$ to the damping integral. This has the advantage of allowing the damping term to be precalculated for a series of values of $\eta$, and then interpolated ready to use in a new iteration of the ODE solution. This makes the solution tractable, even if still somewhat numerically expensive, involving $\mathcal{O}(N)$ integrals for a total $\mathcal{O}(N^2)$ algorithm.

The iteration process ends by some metric of convergence. Here we sum the squared residuals between consecutive solutions for $\Tgw^{\prime\,\,{(N)}}$ and divide by the number of timesteps and the wavenumber $k$. The former division guarantees an intensive metric for convergence -- independent of number of points considered -- and the latter accounts for the derivatives being $\simeq k$ larger than the transfer functions\footnote{Alternatively residuals between undifferentiated $\Tgw$ can be considered, but derivatives are already stored in memory for the damping integral, hence this approach constitutes a memory saving.} which are bounded $-1\leq\Tgw\leq 1$.

Depending on the chosen wavenumber and desired precision, the method takes $\simeq 5$--$10$ iterations to reach a final converged solution. This takes just a few seconds after moderate optimisation, using ODE solvers in the anisotropy module of \cosmotherm \citep{Chluba:2011hw}. This made it possible, with parallelisation, to quickly solve the many tens or hundreds of thousands of $k$ values needed to fit accurate envelopes to energy spectra.
\subsection{Comparison to analytic results}
\label{subsec:analytic_comparison}
\begin{figure}
\centering 
\includegraphics[width=\columnwidth]{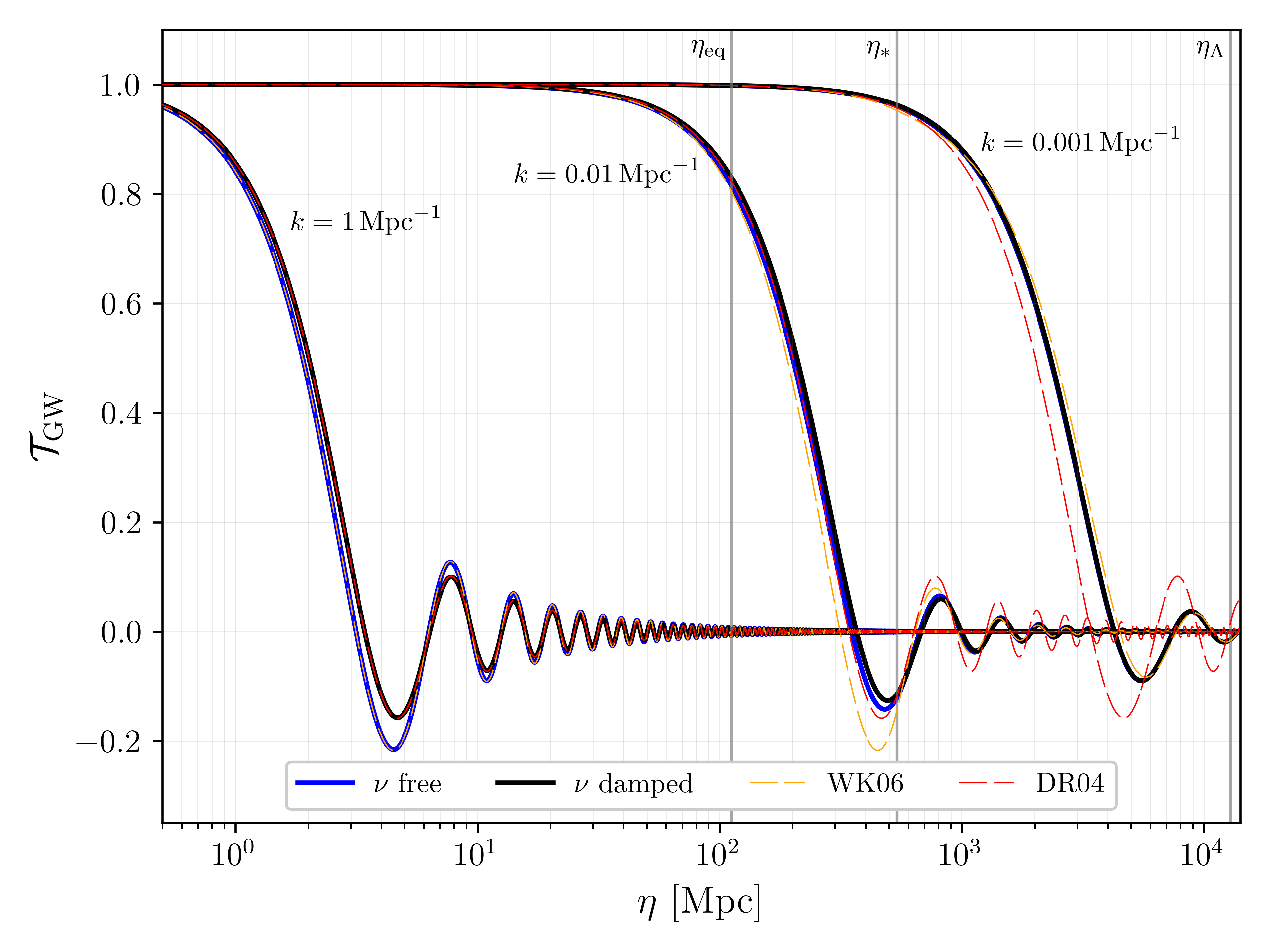}
\caption{A graph showing examples of $\Tgw$ for $k/\Mpc^{-1} \in$ [$1$, $0.01$, $0.001$]. Cases with neutrino damping (black) and without (blue) are shown. The former is approximated by WK06 (orange) and the latter by DR04 (red). It can be seen that the free function is well approximated in each era, but not in the RD--MD transition. The damped function was approximated only in the RD era, hence showing large discrepancies at late times. It should be noted, however, that shortly after the transition the damping becomes negligible.}
\label{fig:GW_transfer_examples}
\end{figure}
\begin{figure}
\includegraphics[width=\columnwidth]{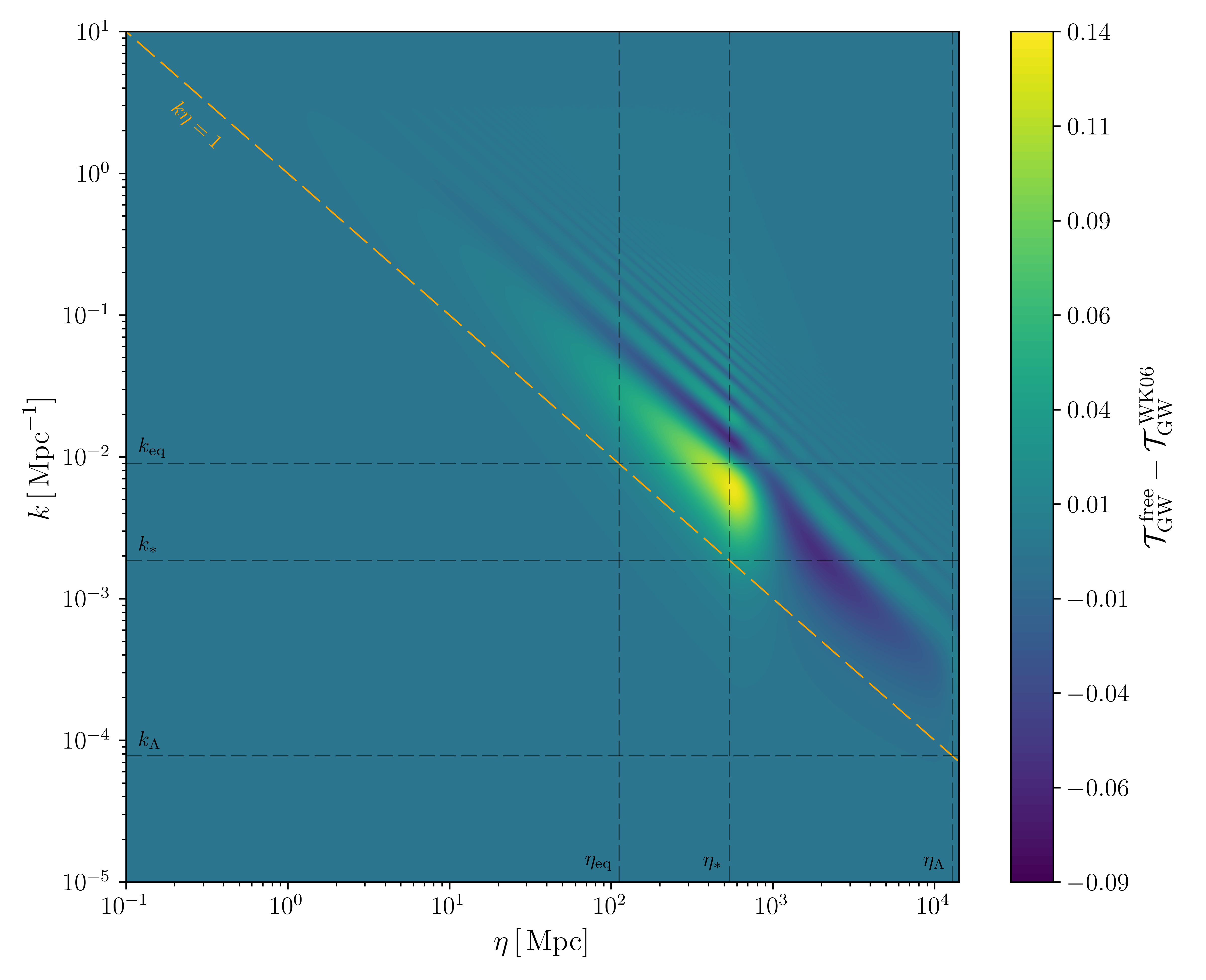}
\caption{Contour plots showing the difference between this work's numerical solutions and approximations given by WK06, against both wavenumber $k$ and conformal time $\eta$. An orange dashed line shows $k\eta=1$. Gray dashed lines show RD--MD transition scales, $k_{\rm eq}$ and $k_*$, as defined in Sect.~\ref{subsec:transfer_functions}}
\label{fig:free_residual_contour}
\end{figure}
\begin{figure}
     \begin{subfigure}[b]{0.5\textwidth}
         \centering
         \includegraphics[width=\textwidth]{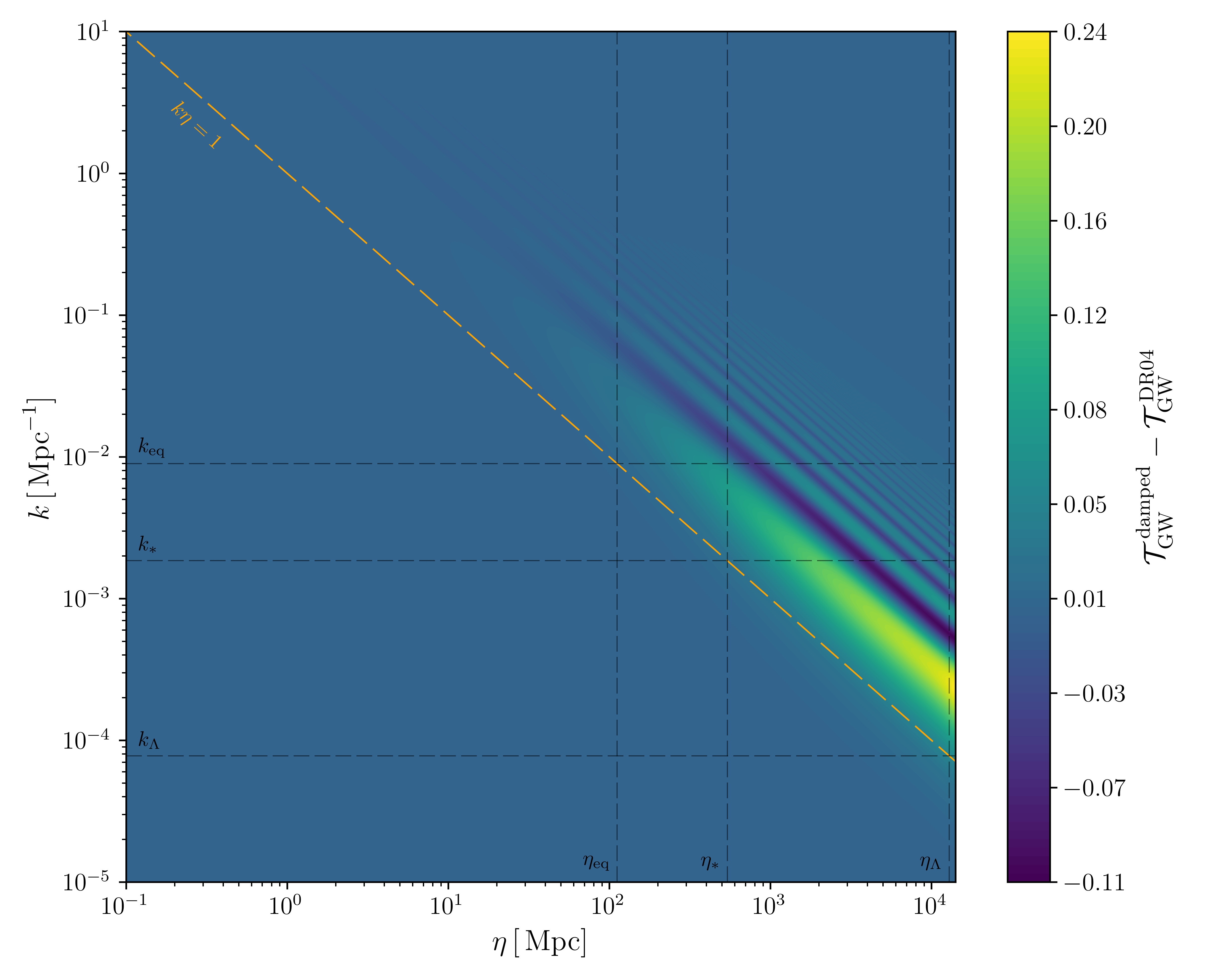}
         \caption{DR04 residuals}
         \label{subfig:damped_residual}
     \end{subfigure}
     \hfill
     \begin{subfigure}[b]{0.5\textwidth}
         \centering
         \includegraphics[width=\textwidth]{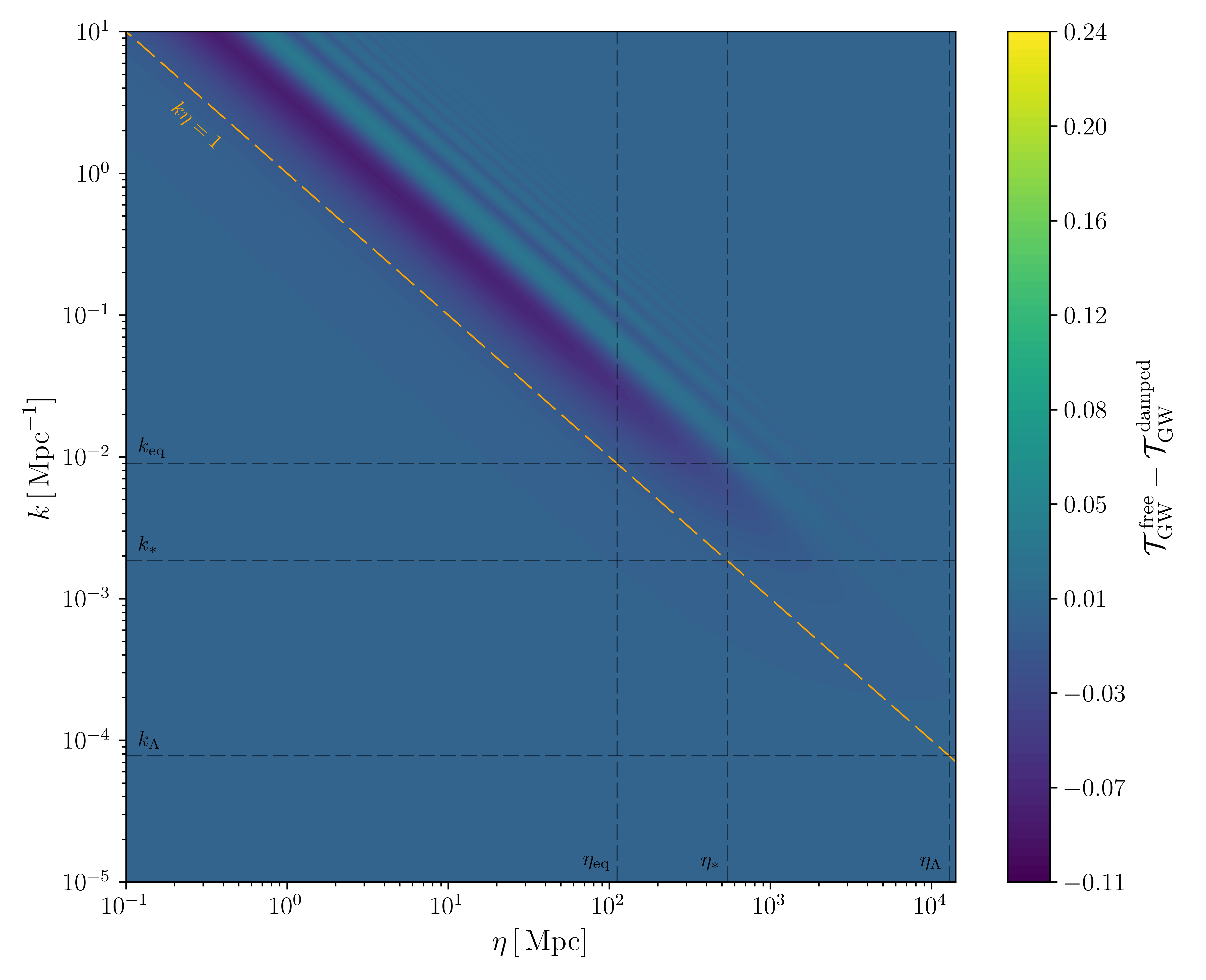}
         \caption{Damping effects}
         \label{subfig:TK_free_damp_comp}
     \end{subfigure}
     \caption{Contour plots showing where the presence of damping is most important (top), and
     differences between this work's numerical solutions and approximations given by DR04 (bottom). This shows that the residuals in the right panel are mostly driven by the end of RD, with only a small intermediate phase showing both MD and damping behaviour simultaneously. An orange dashed line shows $k\eta=1$. Gray dashed lines show RD--MD transition scales, $k_{\rm eq}$ and $k_*$, as defined in Sect.~\ref{subsec:transfer_functions}.}
     \label{fig:damping_comparison}
\end{figure}

To compare the analytic and numerical results we show a qualitative comparison [Fig.~\ref{fig:GW_transfer_examples}], and quantitative comparisons [Figs.~\ref{fig:free_residual_contour},\ref{fig:damping_comparison}]. The former illustrates that the waves usually differ more by an offset in phase than a difference in overall amplitude. With that in mind, we can properly interpret the contour plots, which reveal residuals oscillating throughout the parameter space. This suggests that integrated quantities across either time or wavenumber would be more accurate than these figures initially suggest.

Deep in the RD era, we see an excellent agreement with both WK06 and DR04 as expected. Deep in the MD era on the other hand we see that we again have good agreement with WK06, albeit worse than before due to the matching conditions which essentially provide the MD initial conditions. Exactly at the transition is where the most discrepancy is seen, although the RD--MD transition is relatively short lived [Fig.~\ref{fig:free_residual_contour}]. The approximation by DR04 becomes progressively worse into the MD era, since their coefficients were derived assuming a RD Universe, accounting for the dominant part of the residuals [Fig.~\ref{subfig:damped_residual}]. We also provide Fig.~\ref{subfig:TK_free_damp_comp}, which reveals how quickly the damping ceases once the MD era starts. Through a comparison of the two figures therefore we can see that most of the DR04 residuals are not from a poor modelling of damping, but simply from not capturing the MD dynamics.

In summary these comparisons highlight the robustness of the analytic approximations as well as the utility of the numerical solution by showing the latter can fill the gaps expected from former, but only in specific and brief regimes.

\subsection{Simple fits for the large-scale GWB}
\label{subsec:simple_fits}
\begin{figure}
\centering
\includegraphics[width=\columnwidth]{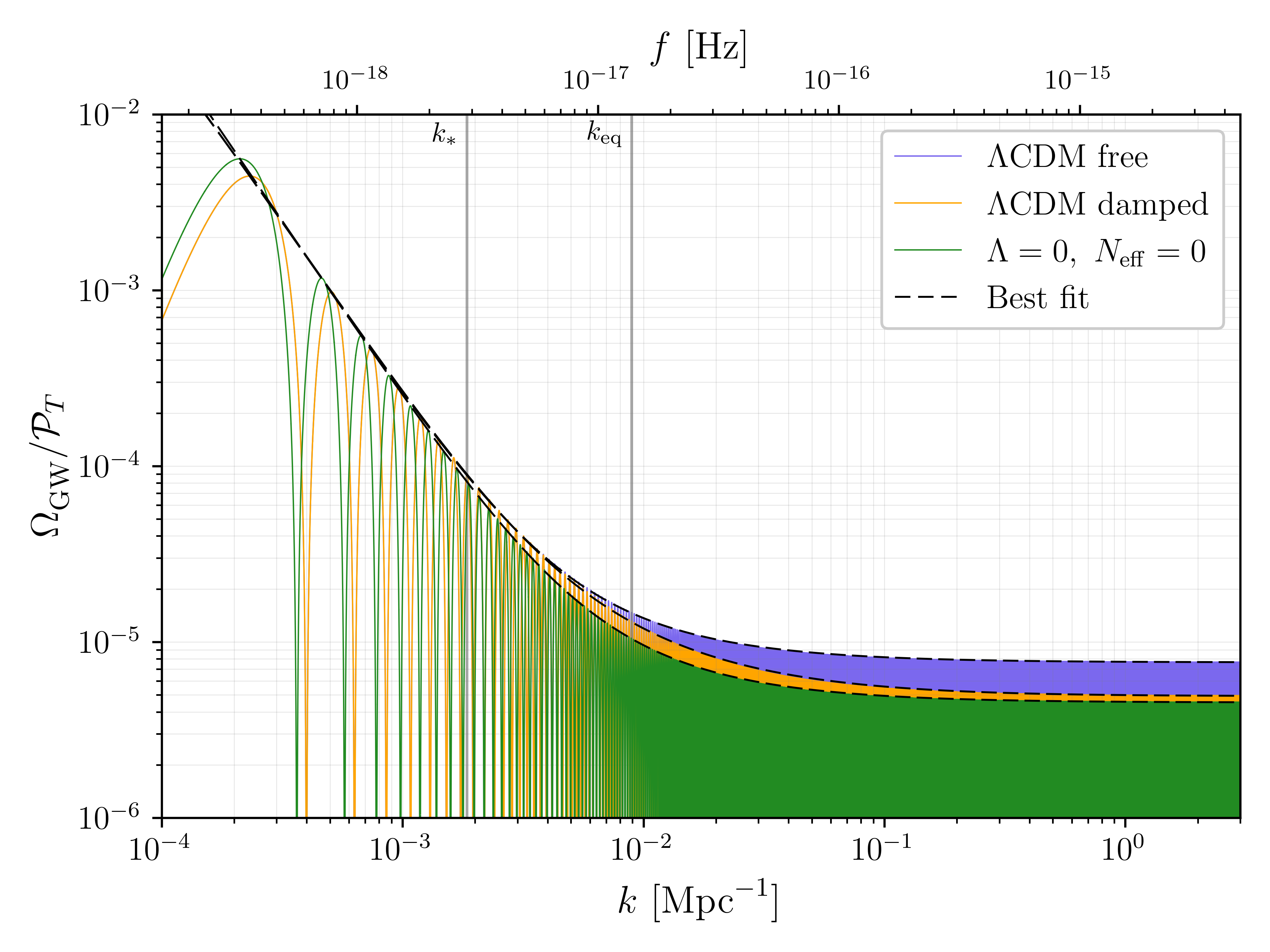}
\caption{A graph showing $\Omega_{\rm GW}/\mathcal{P}_T$ across $k$ as seen at $\eta_0$. Two distinct regions are discernible: a $1/k^2$ slope and a flat branch, corresponding to modes entering horizon during MD and RD, respectively. Spectra are shown for full $\Lambda$CDM with and without neutrino damping, and a simplified cosmology with $\Lambda=0$, $N_{\rm eff}=0$. The best-fit envelopes (i.e., twice the average power) for these cosmologies are shown with black lines, and given in Table~\ref{table:envelope_fits}.}
\label{fig:GW_enevelope_comparison}
\end{figure}
In this subsection, we will give approximate fits for the energy spectrum of the GWB as measured at $\eta_0$ derived from the fully numerical treatment. This approach is greatly facilitated by knowing a sensible functional form in which to package the results, as was discussed in Sect.~\ref{subsec:energy_spectrum}.
Expecting a spectrum that interpolates between $\kappa^{-2}$ and $\kappa^{0}$ motivates a more general formula for approximating the numerical results, which simply includes more general powers to capture the subtleties of the RD--MD transition\footnote{Other attempts included having $\kappa^{-1/2}$, and allowing for a general power $\kappa^\gamma$. The fit used in the main text was chosen through trial and error, showing better results with simpler coefficients than the other functional forms.}. As well as including powers between $-2$ and $0$ we also include an inverse cubic term\footnote{This term was not included in previous work \citep{Kite:2021GW2SD}. The changes however are only a few percent, and outside the scales visible to $\mu$ distortions.}, which specifically in the $\nu$ damped scenario helps with the sharper increase in the spectrum as damping ceases for low $k$:
\begin{equation}
    \bigg\langle\frac{\Omega_{\rm GW}}{\mathcal{P}_T}\bigg\rvert_{\xi_0}\bigg\rangle = \left(\frac{1}{2}\right)\mathcal{D}\,\frac{\omrel}{12}\left(1 + \alpha_1\kappa^{-1} + \alpha_2\kappa^{-3/2} + \alpha_3\kappa^{-2} + \alpha_4\kappa^{-3}\right),
    \label{eq:envelope_func}
\end{equation}
where $\mathcal{D}$ is a coefficient to represent neutrino damping that we discuss more generally in Sect.~\ref{subsec:neutrino_damping}. The quality of this fit can be seen in Fig.~\ref{fig:GW_enevelope_comparison}, with corresponding coefficients given in Table~\ref{table:envelope_fits}. For ease of comparison three fiducial Cosmologies have been chosen: the first two corresponding to a best-fit \textit{Planck} 2018 Universe, with and without neutrino damping, while the third is a simplified cosmology neglecting both neutrinos and $\omlambda$.
This third cosmology highlights how simultaneously neglecting the neutrino contribution to the energy budget and the consequent neutrino damping will coincidentally lead to almost correct results, departing from the full solution by only $\sim6\%$. We hope by providing this fit it may be easier to diagnose oversights in the literature (a similar cancellation of mistakes with almost correct results will be discussed in Sect.~\ref{subsec:late_acceleration}).

By using the natural scale to define $\kappa$ we yield simple values for the $\alpha$ coefficients, with some implicit degree of cosmology independence (Note the similarity between the first and third row in Table~\ref{table:envelope_fits}, once accounting for changes in $\eta_*$).

Although only shown to $k=3\,\Mpc^{-1}$ here, that is sufficient to show the limit of the GWB envelope indeed tends to $\omrel/12$. This is important, as the amplitude of the spectrum can be extrapolated beyond this scale without full calculation: cosmological dependence is imprinted at the time of horizon crossing, and so evaluating a given solution to $k\eta\gtrsim100$ with and without some physical effect allows one to extrapolate\footnote{Note that \cite{saikawa:2018} perform a similar extrapolation on the transfer functions themselves using the WKB approximation.} the spectrum appropriately by multiplying the appropriate ratio of transfer functions by $\omrel/12$. We use this to model the effects of changing relativistic degrees of freedom in Sect.~\ref{subsec:relativistic_dof}, which all occur at scales $k\gtrsim 10^3\,\Mpc^{-1}$.
\begin{table*}
\begin{center}
\begin{tabular}{ l | c c c c c c c }
Cosmology & $\mathcal{D}$ & $\omrel$ & $\eta^*$
& $\alpha_1$ & $\alpha_2$ & $\alpha_3$ & $\alpha_4$ \\
\hline
\lcdm\ free & $1$ & $9.18\times 10^{-5}$ 
& 540.44\,Mpc
& $4.15$ & $-4.55$ & $11.08$ & $-0.11$ \\
\lcdm\ damped & $0.642$ & $9.18\times 10^{-5}$ 
& 540.44\,Mpc
& $8.06$ & $-8.46$ & $17.86$ & $-0.20$ \\
$\Lambda=0$, $N_{\rm eff}=0$ & $1$ & $5.43\times 10^{-5}$ 
& 415.50\,Mpc
& $4.17$ & $-4.21$ & $10.55$ & $-0.01$
\end{tabular}
\caption{Coefficients to calculate $\Omega_{\rm GW}/\mathcal{P}_T$ from Eq.~\eqref{eq:envelope_func} for three fiducial Cosmologies.
}
\label{table:envelope_fits}
\end{center}
\end{table*}

\section{Analysis of GWB features}
\label{sec:GWB_features}
In this section, we discuss some of the GWB features previously mentioned in closer detail, with specific focus on the consequences for SD constraints on the GWB. We start with aspects of neutrino damping, then cover the late dark energy domination and finish with a discussion of the early thermal history.

\subsection{Neutrino damping}
\label{subsec:neutrino_damping}
By comparing the solution to Eq.~\eqref{eq:master_gw_eq} with and without the damping integral, and ignoring phase shifts of the transfer function, we can define the damping factor $\mathcal{D}$ as the ratio of the amplitudes squared: $\left[\Tgw^{\rm damped}\right]^2\approx\mathcal{D}\left[\Tgw^{\rm free}\right]^2$. This definition mirrors that introduced in Sect.~\ref{subsec:simple_fits}, but can now be applied to a single wave of wavenumber $k$. This is useful since the damping will only affect a finite range of scales, essentially giving $\mathcal{D}\equiv \mathcal{D}(k)$.

\begin{figure}
\centering
\includegraphics[width=\columnwidth]{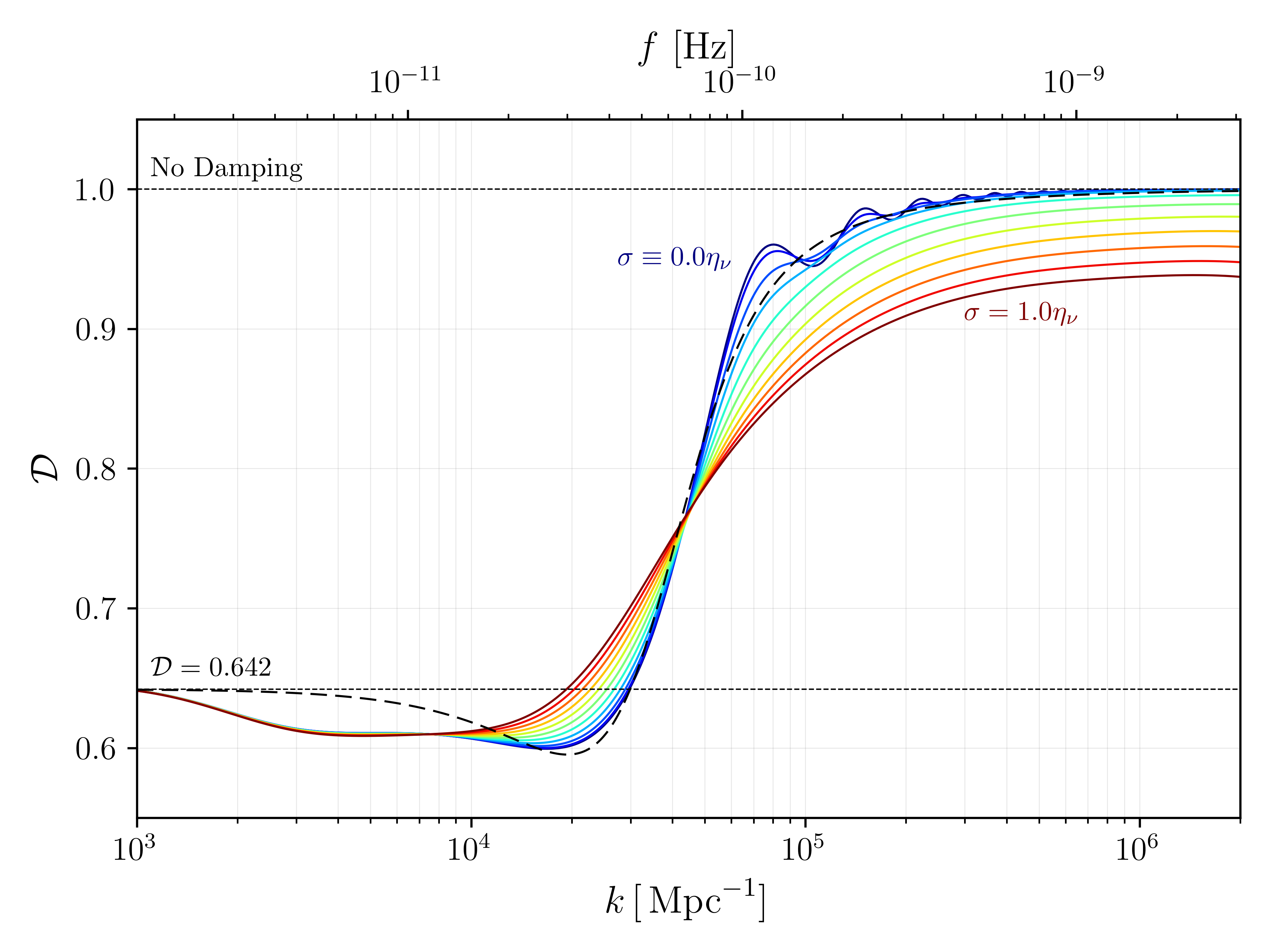}
\caption{Illustration for the damping envelope at large $k$. At $k\gtrsim \pot{3}{5} \, \Mpc^{-1}$, the envelope has essentially returned to the undamped solution, while by $k\simeq 10^3 \, \Mpc^{-1}$ the waves reach the expected $\mathcal{D}=0.642$. A parameter $\sigma$ modulates the {\it sharpness} with which the neutrino decoupling occurs. Each increment in colour corresponds to increasing $\sigma$ by $0.1$. The dashed line shows the simple approximation given in Eq.~\eqref{eq:D_cutoff_approx}.}
\label{fig:damping_width}
\end{figure}

Specifically we expect $\mathcal{D}$ to tend to unity both for large and small $k$ -- the former since the modes were subhorizon before neutrinos started free-streaming, and the latter because the energy density of neutrinos was too small to have considerable effects. The low-$k$ shape of $\mathcal{D}$ is intrinsically linked with the MD--RD transition, a moment which was relatively recent in cosmological history. The exact low-$k$ dependence of $\mathcal{D}$ therefore does not manifest clearly in the GWB as seen today [e.g. convergence of damped and undamped solutions would look differently in Fig.~\ref{fig:GW_enevelope_comparison} if the Universe was older/younger]. Instead we turn our attention to the shape of the damping envelope for large $k$, which in contrast reveals itself clearly, as seen in Fig.~\ref{fig:damping_width} (for $\sigma=0$, as introduced next). The figure shows an expected smooth transition from the previously discussed damping constant and unity, but with a plateau around $\pot{2}{3} \lesssim k/\Mpc^{-1} \lesssim 10^4$. This feature is associated with the energy introduced to the medium from electron-positron annihilation (see Sect.~\ref{subsec:relativistic_dof}), which prolongs the time at which the universe has $T\approx 2\,{\rm MeV}$, the temperature of neutrino decoupling.

The damping integral in Eq.~\eqref{eq:damping_integrand} assumed an instantaneous decoupling of the neutrinos, which leads to oscillations in the damping envelope, as noted in WK06. A more realistic scenario can be achieved by introducing a factor to the integrand which smoothly tends to $0$ for $\eta<\eta_\nu$ and to unity for $\eta>\eta_\nu$, with some characteristic width $\sigma$ governing the sharpness of transition:
\begin{equation}
    \int_{\eta_{\nu}}^{\eta}\,\left(\cdots\right) \,\,\longrightarrow\,\, \int_0^{\eta}\,\left[\frac{1+\tanh{(\frac{\eta-\eta_\nu}{\sigma})}}{2}\right]\left(\cdots\right).
\end{equation}
The effects of this $\sigma$ are also shown in Fig.~\ref{fig:damping_width}. Moving forward we adopt a fiducial value of $\sigma=0.2\eta_\nu$, which quickly converges to the correct limits without spurious oscillations. The curve can be approximately replicated by replacing the factor of $\mathcal{D}$ in the second row of Table~\ref{table:envelope_fits} with
\begin{equation}
\label{eq:D_cutoff_approx}
    \mathcal{D}(\chi) \approx 0.642 + (1-0.642)
    \frac{(11.08\,\chi)^3 - (10.78\,\chi)^2}
    {(11.08\,\chi)^3 - (9.41\,\chi)^2 + 1},
\end{equation}
where $\chi=k/\pot{3.5}{5}$. This approximate curve is shown as a dashed line in Fig.~\ref{fig:damping_width}. We will see in Sect.~\ref{subsec:relativistic_dof} that this can be used to replicate the entire GWB spectrum to arbitrarily large $k$.

Although this modified treatment of neutrino decoupling is by no means considered accurate, it highlights an important dependence of the precise shape of the damping envelope on the decoupling physics. As mentioned above, a more accurate treatment including the full decoupling, neutrino oscillations and possible neutrino spectral distortions should be considered, is, however, beyond the scope of this work.

The ceasing of damping effects at large $k$ has not been included in the calculation of spectral distortion window functions \citep{Chluba2014, Kite:2021GW2SD}, meaning these have been underestimated. With a full calculation we would see a boost in sensitivity of $\simeq 36\%$ on scales $10^5 \lesssim k/\Mpc^{-1} \lesssim 10^8$. This nearly corresponds to a factor of two in the observing time, rendering this correction non-trivial. However, at $k\gtrsim 10^8\,{\rm Mpc}^{-1}$, the effect of relativistic degrees of freedom become more important, almost exactly canceling this omission (again, see Sect.~\ref{subsec:relativistic_dof}).
\subsubsection{Cosmology dependence of the damping coefficient}
The total amplitude of the damping carries Cosmological dependence in the form of $f_{\nu,0}=\Omega_\nu/(\Omega_\gamma+\Omega_\nu)$, as defined in Eq.~\eqref{eq:damping_integrand}, which in turn will depend on $N_{\rm eff}$. Using the iterative procedure for the damping contributions (see Sect.~\ref{sec:numerical_solutions}) we find a \lcdm\ value of $\mathcal{D}=0.642$, differing slightly from \cite{Weinberg:2004:nuDamp}, where it was concluded that $\mathcal{D}=0.644$ by using $N_{\rm eff}=3$, implying $f_{\nu,0}=0.40523$. By running the solution for values $0\leq f_{\nu,0} \leq 1$ we find the fit
\begin{equation}
    \mathcal{D}= 1 - 0.45397 \,\zeta + 0.11375 \,\zeta^2 - 0.01904 \,\zeta^3 + 0.00168 \,\zeta^4
\end{equation}
where $\zeta=f_{\nu,0}/0.40890$. This choice of pivot value is derived from the theoretically expected $N_{\rm eff}=3.046$, since the current measured value is poorly constrained to $N_{\rm eff}=2.99\pm 0.17$ \citep{Planck2018params}.
Inserting $f_{\nu,0}=0.40523$, we obtain $\mathcal{D}=0.6449$, which more closely matches the value of  \cite{Weinberg:2004:nuDamp}, and matches the $\mathcal{D}=0.645$ of DR04, based on the same $f_{\nu,0}$, thus confirming the equivalence of our treatments.

\subsection{Late time acceleration}
\label{subsec:late_acceleration}
\begin{figure}
\centering
\includegraphics[width=\columnwidth]{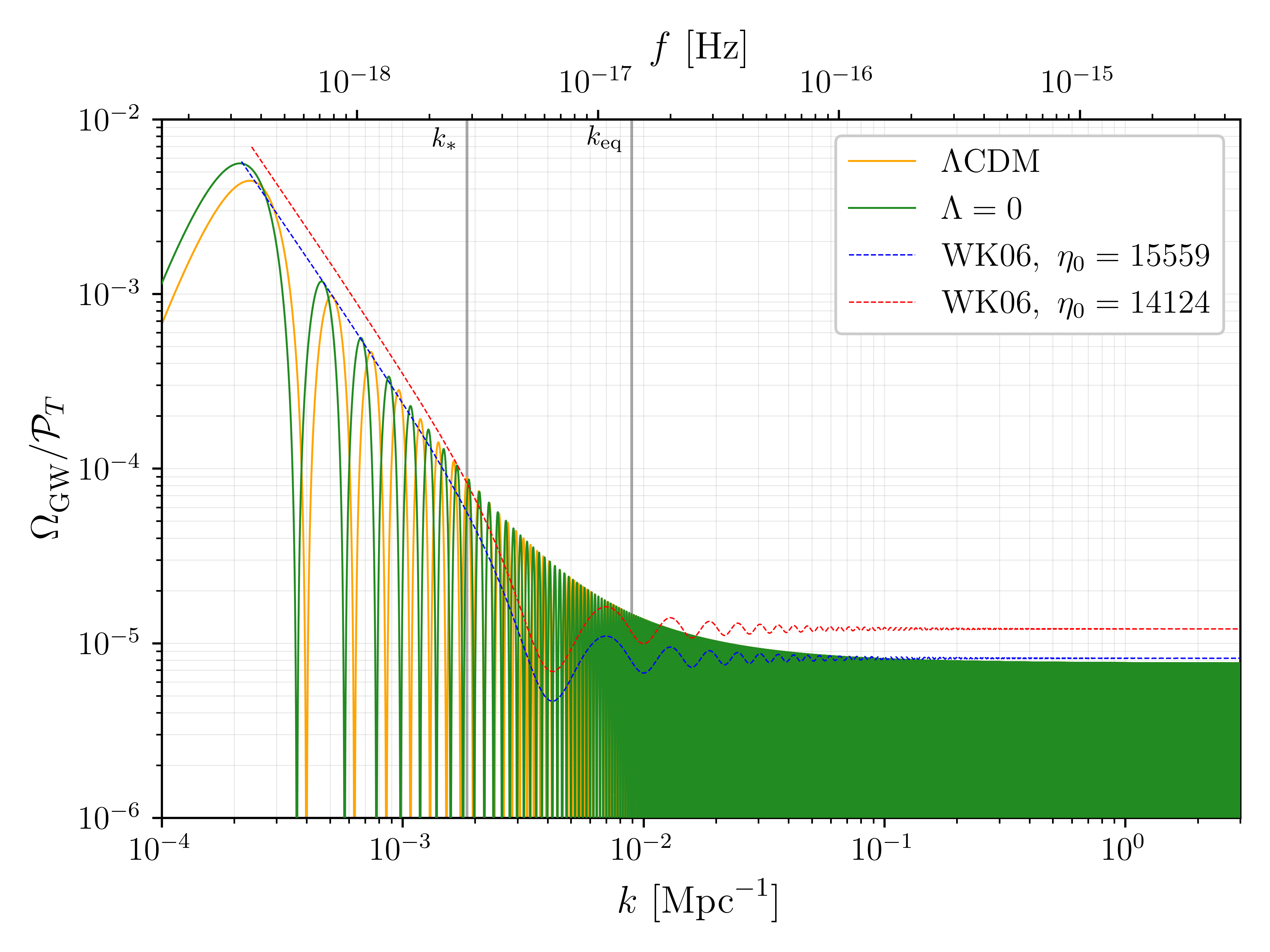}
\caption{A graph showing $\Omega_{\rm GW}/\mathcal{P}_T$ across $k$ as seen at $\eta_0$ in two different Universes: a standard \lcdm\ and one without a cosmological constant. Differences are only present for the smallest $k$ values, and correspond to phase shifts rather than a difference in fundamental spectral shape.}
\label{fig:GW_enevelope_nolam_comparison}
\end{figure}
The inclusion of a DE component in the Universe's expansion does not allow for simple analytic expressions like those given in Eq.~\eqref{eq:MD-RD-expansion}. Despite this, we can argue that the effects of a cosmological constant will be small, since this component only becomes dominant at very late times. The scale factor at which $\omlambda$ matches the contribution from $\ommat$ is given by
\begin{equation}
    a_{\Lambda} = \left(\frac{\ommat}{\omlambda}\right)^{1/3},
\end{equation}
which takes a value of $a_\Lambda\simeq 3/4$ with current best-fit parameters \citep{Planck2018params}, a value close to today's scale factor $a_0=1$ (e.g. see vertical lines in Fig.~\ref{fig:GW_transfer_examples}). Recalling that the effects of cosmological expansion are imprinted on the GWB at the time of horizon crossing, this means that only large scales which crossed horizon recently can be impacted in spectral shape. This can be verified with the numerical solution -- where arbitrary expansion histories are easily included -- as can be seen in Fig.~\ref{fig:GW_enevelope_nolam_comparison}. We see that the spectral shape is unchanged, with only specific local maxima showing a shift of position, and most of the spectrum simply receiving a phase shift. Sensitivity to the shifted peaks would require sensitivity to wavelengths spanning large fractions of the observable Universe, which even if feasible would be heavily limited by cosmic variance. Similarly the phase shift is invisible to probes which typically average the spectrum over one or many cycles. This suggests therefore that late acceleration can be neglected for practical purposes.

Although the shape of the spectrum does not change significantly, one important effect is in reducing the value of $\eta_0$. In words, if one includes late time accelerated expansion then the waves have less time to evolve before the scale factor reaches today's value of $a_0~=~1$. This means that when using the approximations given in Eq.~\eqref{eq:free_transfer} one should use the \textit{wrong} value $\eta_0\approx15560\,\Mpc$ for more accurate results in a full \lcdm\ Universe. Using the \textit{correct} value of $\eta_0\approx14120\,\Mpc$ gives a correspondingly younger spectrum, and hence overall larger amplitude. This is shown in Fig.~\ref{fig:GW_enevelope_nolam_comparison}, where we depict the expected limits of the analytic solution by plotting an interpolated line of the local peaks.

Another potential cancellation of errors arises here. By inspecting the second line in Eq.~\eqref{eq:energy_high_kappa} we see that, with $H_0$ held constant, the fundamental dependence of the energy spectrum is $\propto \eta_*^2/\eta_0^4$. A cancellation of errors, which in fact gives the correct result to within $\simeq 15\%$, is to use the lower $\eta_*$ from neglecting neutrinos, with the lower $\eta_0$ from a younger late-accelerated universe. This again makes diagnosing discrepancies in the literature difficult, especially if there is ambiguity between $\omrel$ and $\Omega_\gamma$.
\subsection{Relativistic degrees of freedom}
\label{subsec:relativistic_dof}
\begin{figure*}
\centering
\includegraphics[width=0.96\linewidth]{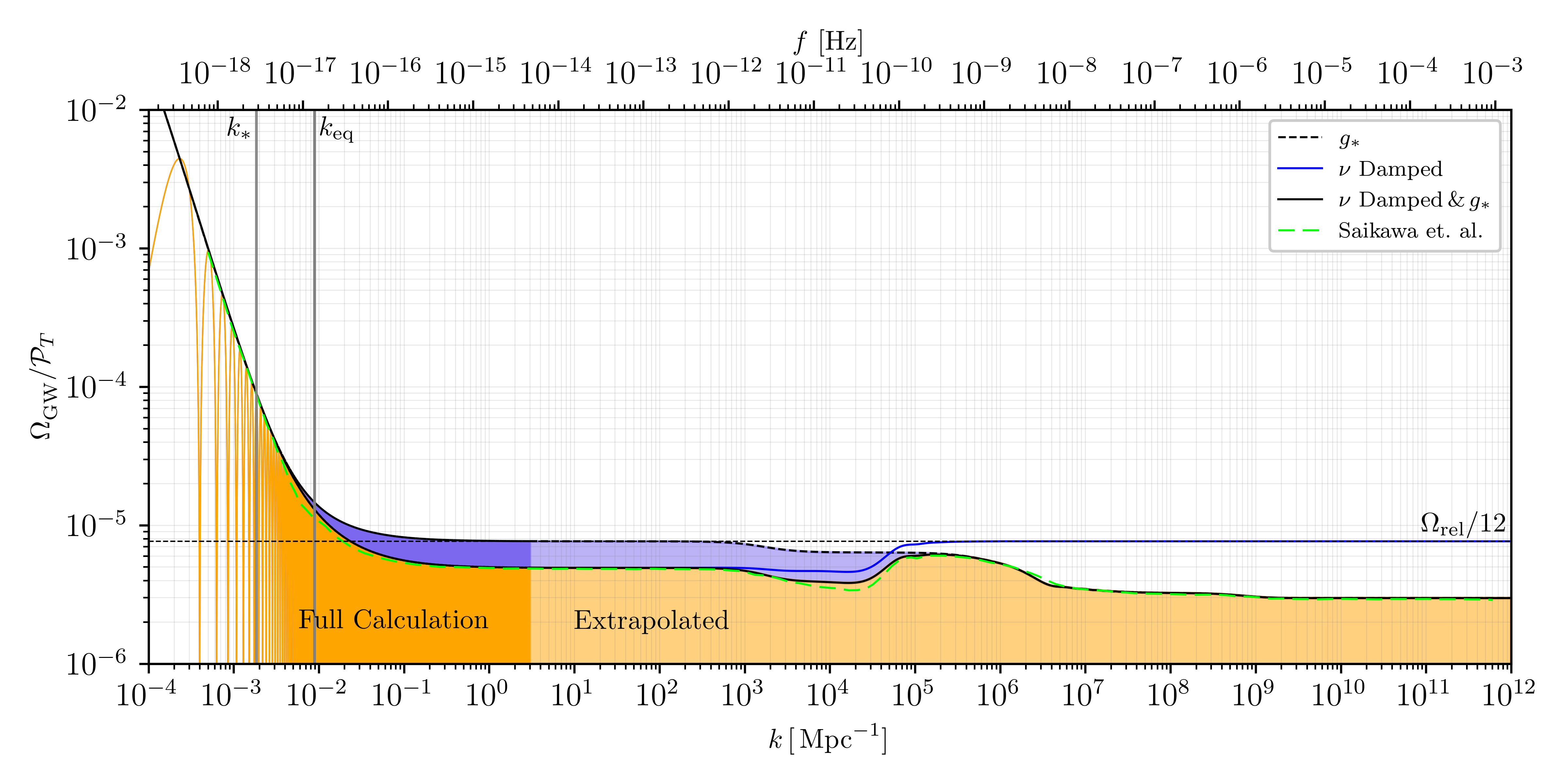}
\caption{A graph showing the GWB for an extremely wide range of scales, extending to the highest frequencies interferometry missions aim to measure, and also the highest energy scales that known physics allow us to model. We show the onset of the effects of damping and the effects of changing relativistic degrees of freedom. For comparison we show digitalised data from \protect\cite{saikawa:2018}, which matches well to the findings here.}
\label{fig:GW_gstar}
\end{figure*}
Accounting for the changing relativistic degrees of freedom involves modifying the evolution of the scale factor \citep[e.g.,][see]{Watanabe2006}:
\begin{equation}
a'=a^2 H_0 \sqrt{
\frac{g_{*\rho}}{g_{*\rho 0}}\left(\frac{g_{*s0}}{g_{*s}}\right)^{4/3}\omrel a^{-4}
+ \ommat a^{-3} + \omlambda}.
\end{equation}
The functions for $g_{*\rho}$ and $g_{*s}$ are available in pretabulated or functional forms, together with much more detailed discussion of the physics at play, in \cite{saikawa:2018}.

This change in the energy budget can be interpreted as a departure from the expected $\rho \propto a^{-4}$ behaviour of relativistic fluids, but only in specific temperature ranges where there is some change in the thermodynamics of the plasma, e.g., during phase transitions. These small changes in the expansion rate will be imprinted on the GWB, as illustrated in Fig.~\ref{fig:GW_gstar}, where we again numerically solved the transfer functions with modified expansion rates. It is noteworthy that this change in the energy budget of relativistic particles changes the exact relation between conformal time $\eta$, scale factor $a$, and temperature $T$.

Fortunately, the lowest frequencies impacted by changes in the relativistic energy budget are $k\simeq 10^3\,\Mpc^{-1}$, which is deep into the region of tensor modes which entered horizon during RD. This means the limiting case of $\langle\omgw\rangle=\left(1/2\right)\omrel/12$ in the absence of any other physical effects can be safely extrapolated (see discussion at the end of Sect.~\ref{subsec:simple_fits}). In particular an extra factor which approximately accounts for the changes in the relativistic degrees of freedom is given by \citep{saikawa:2018}
\begin{equation}
    \langle\omgw^{g_*}\rangle \approx \langle\omgw\rangle
    \left(\frac{g_\rho}{g_{\rho 0}}\right)
    \left(\frac{g_{s0}}{g_{s}}\right)^{4/3}.
\end{equation}
Using the \lcdm\ envelopes given in Table~\ref{table:envelope_fits}, noting that the $\nu$ damped envelope requires the modification in Eq.~\eqref{eq:D_cutoff_approx}, matches the full solution to within $\simeq 5\%$ as measured with respect to our full numerical solution. This means that the relatively simple fits in this paper together with the pretabulated $g_*$ functions (see specifically Appendix A in \cite{saikawa:2018}) the entire \lcdm\ GWB can be replicated to high precision across a very large range of scales.

The $g_*$ effects discussed here were again neglected in previous calculations of the spectral distortion window functions\footnote{Note however that some of the models discussed in \citet{Kite:2021GW2SD} implicitly included the $g_*$ effects in their energy spectra, and were thus indirectly included in the distortion calculation. In future these should be included in the window function itself for completeness and higher accuracy.}. If the effects were included they would remove a similar percentage of sensitivity as the damping effects, but over a different range of scales. This combined with the over-extension of damping effects will lead to almost unchanged results in \cite{Kite:2021GW2SD}, but should be fully accounted for in future distortion studies.

\section{Discussion and conclusion}
\label{sec:conclusion}
The GWB offers an exciting new window to the physics of the early Universe, and a diverse set of probes will soon begin the search for this new signal. In this paper we give simple yet precise large-scale functional forms for the GWB energy density, which can aid in estimating the efficacy of some observations. The scales considered here are especially helpful in comparing CMB $B$-modes and CMB spectral distortion measurements \citep[as required in][]{Kite:2021GW2SD}. More generally, however, any comparison between early- or late-universe probes requires understanding of the $\eta_*$ scale for MD--RD transition.

In this paper we endeavoured to firstly elucidate the physics at play in the GWB in a pedagogical way, secondly to provide tools for simple calculation of the large-scale energy spectrum, and finally discuss the main features in the GWB with special attention on consequences for spectral distortion calculations.

In Sect.~\ref{sec:GWB_physics} we qualitatively review the physical phenomena affecting the large-scale spectrum. The most important of these is the transition between a universe dominated by relativistic particle species and matter. We explicitly discuss the importance of including neutrinos, despite the common misnomer of \textit{radiation domination}. A second important effect we discuss is that of neutrino damping via anisotropic stress in the medium.
In Sect.~\ref{sec:analytic_solutions} we give solutions to the transfer function valid in RD and MD respectively [see Eqs.~\eqref{eq:free_transfer},\eqref{eq:free_transfer_prime}], and explain how these are matched at the transition assuming this to be instantaneous [Eq.~\eqref{eq:normalisation}]. These solutions give expected limits for the energy density of the GWB [Eqs.~\eqref{eq:energy_high_kappa},\eqref{eq:energy_low_kappa}].
In Sect.~\ref{sec:numerical_solutions} we explain how the damping can be treated through an iterative numerical method. This method is found to give results matching those of WK06 and DR04 in the appropriate limits (see Fig.~\ref{fig:GW_transfer_examples}). Energy spectra as seen today are shown for various limiting cases: with and without damping, with and without DE, with and without neutrinos (see Figs.~\ref{fig:GW_enevelope_comparison},\ref{fig:GW_enevelope_nolam_comparison}). To replicate these spectra with ease, we provide coefficients in Table~\ref{table:envelope_fits} for use with Eq.~\eqref{eq:envelope_func}, which is valid to scales of $k\simeq 10^3\,\Mpc^{-1}$.
Sect.~\ref{sec:GWB_features} finalises the analysis with discussion of various GWB features: cosmological dependence of neutrino damping, effects of late time accelerated expansion, and changes in the number of relativistic degrees of freedom.

The effects included in this analysis were purely standard model Physics. More generally one would apply the techniques discussed here to verify the avenues of discovery for new Physics hidden in the GWB. We note that the numerical method utilized in this work can be straightforwardly generalized to non-standard thermal histories that transition from RD to MD and back one or more times, as is the case in a variety of scenarios of beyond the standard model physics \citep{Acharya:2008bk, Acharya:2019pas, Arbey:2020yzj}. It can also be extended to include the presence of other light, weakly interacting particles, such as axions or axion-like particles in the early universe \citep{Marsh:2015xka}. We defer the implementation of this to a future study.

Two of the features explored in this paper reveal inaccuracies in previous calculations of tensor window functions, $W_\mu(k)$, used to calculate SD amplitudes arising from primordial tensor power spectra \citep[e.g.,][]{Chluba2015}. Previously damping has been included, but extending to arbitrarily high $k$. The damping ceases to affect the spectrum beyond $k\simeq 10^5\,\Mpc^{-1}$ (see Fig.~\ref{fig:damping_width}). However, the effects of the relativistic degrees of freedom were also not included explicitly within $W_\mu$, and would lead to consecutive over- and under-estimations on scales $k \gtrsim 10^3\,\Mpc^{-1}$. Together all these changes add to only small percent changes, rendering the conclusions of \cite{Kite:2021GW2SD} still valid. The calculation of new and more precise window functions remains as future work, where it would be appropriate to include the tensor perturbations within a full Boltzmann code, and accurately model the GW-photon interaction, and thus fully capturing the smooth decoupling of the photons, even in the post-recombination era.

\section*{Data Availability}
Data in all figures available at \url{https://doi.org/10.5281/zenodo.5141789}. A full release of the \cosmotherm code is planned for the near future, including the GW module used here.

{\small
\section*{Acknowledgments}
We would like to thank Eiichiro Komatsu and Yuki Watanabe for valuable comments on the draft.
We also thank Paolo Campeti for helpful discussions on the role of $\eta_0$ on the approximate spectrum.
Finally we thank Chiara Caprini and Daniel Figueroa for their encouraging words on the draft.

This work was supported by the ERC Consolidator Grant {\it CMBSPEC} (No.~725456) as part of the European Union's Horizon 2020 research and innovation program.
TK was further supported by STFC grant ST/T506291/1.
JC was also supported by the Royal Society as a Royal Society URF at the University of Manchester, UK.
}

{\small
\bibliographystyle{mn2e}
\bibliography{bibliografia,Lit}
}

\end{document}